\documentclass[aps,prb]{revtex4}
\usepackage{bm}
\usepackage{amsmath}
\usepackage{color}
\usepackage{graphicx}
\usepackage{amssymb}
\DeclareMathAlphabet{\mathbbmsl}{U}{bbm}{m}{sl}
\makeatletter
\newsavebox{\@brx}
\newcommand{\llangle}[1][]{\savebox{\@brx}{\(\m@th{#1\langle}\)}%
	\mathopen{\copy\@brx\kern-0.5\wd\@brx\usebox{\@brx}}}
\newcommand{\rrangle}[1][]{\savebox{\@brx}{\(\m@th{#1\rangle}\)}%
	\mathclose{\copy\@brx\kern-0.5\wd\@brx\usebox{\@brx}}}
\makeatother
\begin{document}
\draft

\title{Plasmon damping rates in Coulomb coupled 2D layers in a heterostructure}

\author{Dipendra Dahal$^1$, Godfrey Gumbs$^2$, Andrii Iurov$^3$ and Chin-Sen Ting$^1$}

\affiliation{$^1$Texas Center for Superconductivity and Department of Physics,
University of Houston, Houston, TX 77204, USA\\
$^2$ Department of Physics and Astronomy, Hunter College of the City University of New York, 695 Park Avenue, New York, New York 10065, USA\\
$^{3}$Department of Physics and Computer Science, Medgar Evers College of City University of New York, Brooklyn, NY 11225, USA  }

\date{\today}

\begin{abstract}

The Coulomb excitations of charge density oscillation are calculated for a double layer heterostructure.  Specifically, we consider two-dimensional (2D) layers of silicene and graphene  on a  substrate.  From the obtained surface response function, we calculated the plasmon dispersion relations which demonstrate the way in which the Coulomb coupling renormalizes the plasmon frequencies. Additionally, we present a novel result for the damping rates of the plasmons in this Coulomb coupled heterostructure  and compare these results as the separation between layers is varied.   

\end{abstract}

\medskip

\vskip 0.2in


\medskip
\par
\maketitle


\section{Introduction}
\label{sec1}

\begin{figure}
\centering
\includegraphics[width=.65\textwidth]{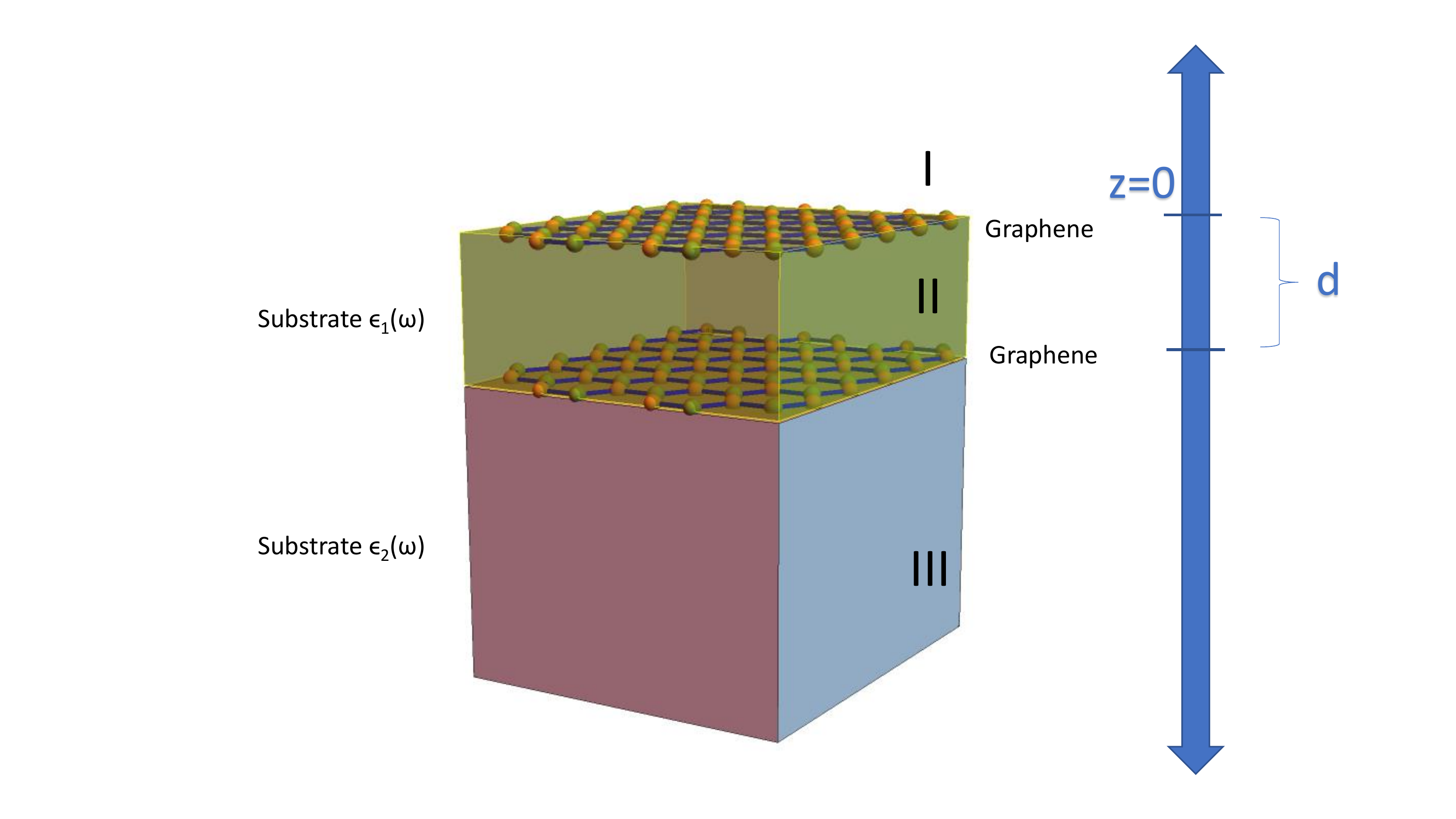}
\caption{(Color online)  Schematic illustration of a  heterostructure consisting of a pair of 2D layers separated by a a dielectric medium $\epsilon_1(\omega)$. This structure lies on a substrate with dielectric function, $\epsilon_2(\omega)$}.
\label{FIG:1}
\end{figure}

A huge number of researchers from various disciplines have been showing their interest in the new material silicene especially, after the development of its fabrication process in 2012. \cite{Vogt}  Because of its exceptional potential applications in electronic and optoelectronic devices, many industries are making substantial investments  to harness its properties. Additionally, before making investments for commercial gain,    both theoreticians and experimentalists have been exploring this material for many years. A credit of  foremost importance goes to Takeda and Shiraishi,\cite{Takeda} who, in 1994, dealt with the atomic and electronic structure of the material for the first time.  These authors  calculated the band structure of silicon in the corrugated stage having optimized atomic geometry. This work, though very novel, did not receive the attention it deserves until in 2004  when single-layer carbon atoms named graphene were fabricated in the laboratory from graphite by Novoselov et al.\cite{novoselov}  Their research not only validated the stability of two-dimensional (2D) material but also opened the door for new research on thin film materials, silicene being one of them.

\medskip
\par

Both silicene and graphene were studied in parallel. The former has a buckled crystal geometry whereas the latter has the honeycomb planar geometry. Due to this, differences arise between them. Ab initio calculations showed that the bandgap of silicene is electrically tunable\cite{Ni,Drummond, Liu} which is an advantageous   property for designing a field effect transistor which works at room temperature. Another distinct difference between these two materials is the strength of the spin-orbital coupling (SOC), which is very weak in graphene. Consequently,   the quantum spin Hall effect occurs at extremely low temperature\cite{kane, Yao}. In contrast to this, silicene displays quantum spin Hall effect at temperature $18$K, far higher than that for graphene. 

\medskip
\par

Several investigations have been carried out on both graphene and silicene with respect to transport phenomena\cite{Bao, Dora, Tian, Rosenstein, Dahal1, Wakamura11, Kim11, Yan11, Bao11}, as well as their magnetic and electric field effects \cite{Gumbs1, Gumbs11, Checkelsky, Nakamura, Yan, Yu, Novoselov, Liu2}, the fabrication process\cite{Eletskii,Jia,Cai,Aliofkhazraei}, and on plasmonic behavior \cite{Garcia, Farmer, Gumbs3, Andrii1, Andrii2, Andrii3, Gumbs4, Dahal, Andrii11}. An intensive literature search on the plasmonic studies suggests  that  no study   on the plasmon dispersion and its rate of damping was carried out for composite silicene and graphene materials. This hybrid material could have  significant benefits for use in  the advancement of: quantum information technology, \cite{Calafell,Alonso,Hanson, Christensen} sensing devices, \cite{Esfandiari,Tong,Hu} and protein analytic  clinical devices, \cite{Hu,Andoy,Viswanathan, Huang} etc.  With this motivation, in this research work, we choose a system composed of silicene and graphene accompanied by a conducting substrates.

\medskip
\par

The plasmon mode is tunable by the thickness of the substrate and the variation of material behavior. We first determine the surface response function of the structure, the same technique used recently by Gumbs et. al.\cite{Gumbs2,Hwang}  which gives us the condition for the existence of the plasmon dispersion. The analytical result for the surface response function is further used for different limiting cases and a comprehensive comparison is made with a variety of structures composed of different graphene-silicene compositions. Furthermore, the same function is used to obtain the Landau damping rate of the plasmon modes whose numerical calculation demonstrates that the it's variation depends on the layer separation, types of dielectric used and the type of 2D layer employed. 

\medskip
\par

 We have organized the rest of our paper as follows. In Sec.\  \ref{sec2}, we present the core idea about our work where we show the analytical result for the surface response function for the chosen structure. Under limiting conditions, the result is used to derive the results for a variety of conditions. The graphical results and their interpretation   are presented in  Sec.\  \ref{sec3}. We conclude our paper with a summary of our main results and conclusions in  Sec.\  \ref{sec4}.

 \section{theory}
\label{sec2}
In this article, we have reported the plasmonic behavior of heterostructure consisting of graphene and silicene together for which the Hamiltonian in the low energy regime near the K point is considered. One significant difference between the Hamiltonian of graphene and silicene is: a small band gap, $\Delta$ is present in the Silicene energy band structure which is due to spin-orbit coupling and applied external electric field. This band gap is not seen in intrinsic graphene.

\subsection{Silicene}

We now briefly describe the case pertaining to silicene whose Hamiltonian in the continuum limit is given by

\begin{equation}
    H_\xi=\hbar v_F(\xi k_x \hat \tau_x+k_y \hat \tau_y)-\xi \Delta_{so}\sigma_z \tau_z+\Delta_z\hat \tau_z \  ,  
\label{eq1}
\end{equation}
where $\hat \tau_{x,y,z}$ and $\sigma_{x,y,z}$ are Pauli matrices corresponding to two spin and coordinate sub-spaces, $\xi=\pm 1$ for the $K$ and $K^\prime$ valleys, $v_F(\approx 5 \times 10^5)$ m/sec is the Fermi velocity for silicene \cite{Ezawa,Drummond}, $k_x$ and $k_y$ are the wave vector components measured relative to the K points. The first term represents the low-energy Hamiltonian whereas the second term denotes the Kane-Mele system \cite{Kane} for intrinsic spin-orbit coupling with an associated spin-orbit band gap of $2\Delta_{so}$.  The last term in the expression describes the sublattice potential difference that arises from the application of a perpendicular electric field.  Equation  (\ref{eq1}) for the Hamiltonian is a block diagonal in $2\times 2$ matrices labeled by valley $(\xi)$ and spin $\sigma=\pm 1$ for up and down spin, respectively. These matrices are given by

\begin{equation}
   \hat H_{\sigma \xi}= \bigg(\begin{array}{cc}
        -\sigma \xi \Delta_{so}+\Delta_z & \hbar v_F(\xi k_x-ik_y)  \\
        \hbar v_F(\xi k_x+ik_y) & \sigma \xi \Delta_{so}-\Delta_z 
    \end{array}\bigg) \   .
    \label{hamiltonian}
\end{equation}
This gives the low-energy eigenvalues as

 \begin{equation}
      E_k=\pm \sqrt{\hbar^2v_f^2|k|^2+\Delta_{\xi \sigma}^2}
     \label{dispersion}
 \end{equation}
 where $\Delta_{\sigma \xi}=|\sigma \xi \Delta_{so}-\Delta_z|$.
 
\subsection{Graphene}
 
 The low-energy model Hamiltonian for  monolayer graphene is similar to that in Eq.\ (\ref{hamiltonian}) with the diagonal terms replaced by zero and  $\xi$  labeling the valley. In this regime, the Hamiltonian for intrinsic graphene  is given by

 \begin{equation}
     \hat H=\hbar v_F\bigg(\begin{array}{cc}
        0 & (\xi k_x-ik_y)  \\
        (\xi k_x+ik_y) & 0
    \end{array}\bigg)
    \label{hamiltonian2}
 \end{equation}
  with the linear energy dispersion,  $ E_k=\pm \hbar v_f |k|$ in either valley.

  \subsection{Polarization function: $\Pi(q,\omega)$}

Considerable work has been done on the dynamical properties involving the use of the dielectric function $\epsilon(q,\omega)$ of various types of free-standing 2D systems\cite{Wunsch, Balassis, Roldan} under different conditions. These include  temperature effects,\cite{Patel,Ramezanali} the role of an ambient magnetic field for the 2D electron gas (2DEG), graphene, silicene,\cite{Tabert} and the dice lattice\cite{Balassis}. For a single 2D layer, one can extract the plasmon dispersion relation and damping rate by employing the dielectric function. However, the situation is more complicated for a multi-layer heterostructure which relies on a knowledge of the surface response function that we have presented in detail below. However, in either  case, we need to calculate the polarization function obtained in the random phase approximation (RPA).  For a 2D layer surrounded by a medium with dielectric constant $\epsilon_b$, the dynamic dielectric function  is given by  

\begin{equation}
      \epsilon(q,\omega)=1-V(q)\Pi(q,\omega) \label{dielfn},
  \end{equation}
where $V(q)=\frac{2\pi e^2}{ 4\pi \epsilon_0  \epsilon_b q}$ is the Coulomb interaction potential and $\epsilon_0$ is the permittivity of free space,  $q$ is the wave vector and $e$ is the electron  charge. The polarization function is an important quantity in calculations of the transport, collective charge motion and, charge screening properties of the material. In the one-loop approximation, the polarization function for gapped graphene is given by \cite{Pyatkovskiy}
	
\begin{eqnarray}
     \Pi^0(q,\omega) &=&  \int  \frac{d^2{\bf k}}{2\pi^2}
\sum\limits_{s,s'=\pm 1}  \left\{  \frac{\hbar^2 v_f^2(\bf k+\bf q) \cdot \bf k+\Delta^2_{\sigma,\xi}}{E_k \cdot E_{|k+q|}}\right\}  \frac{f_0(sE_k-E_F,T)-f_0(s' E_{|k+q|}-E_F,T)}{sE_k-s'E_{|k+q|}-\hbar(\omega+i\delta)} \     ,
\nonumber\\
\label{eq6}
 \end{eqnarray}
where $\theta_{k,k+q} $ is the angle between ${\bf k}$ and ${\bf k}$+${\bf q}$. At zero temperature, the Fermi function $f_0(z)$ is just a step function. The analytical expression for the polarization function for silicene and graphene monolayer is given by Tabert et. al. \cite{Tabert} and Wunsch et. al.\cite{Wunsch} respectively.

\medskip
\par
We now turn our attention to a crucial consideration in this paper regarding the structure consisting of a silicene layer, a graphene layer and  substrates as depicted in Fig. \ref{FIG:1}.    By employing the boundary condition of continuity of the  potential and the discontinuity of the electric field at the interface, we solved for the various coefficients appearing in the potential. The result for the surface response function  $g(q,\omega)$ gives the required conditions for the plasmon dispersion for our case, namely

\begin{eqnarray}
&& \phi_<(z)=e^{-qz}-g(q,\omega)e^{-qz}\  ,   z \lesssim 0 \, , \\
\nonumber
&& \phi_{>}(z)=a_1 e^{-q z}+b_1 e^{q z} \  ,  0 \leq z \leq d \, , \\
\nonumber 
&& \phi_{1>}(z)=k_1e^{-qz} \   ,  z\geq  d \, . 
\end{eqnarray}

Here, $\phi_<(z)$, $\phi_>(z)$ and $\phi_{1>}(z)$ correspond to the electrostatic potential of region (I),(II), and (III) respectively as shown in figure \ref{FIG:1}. In order to conduct numerical computation, we make use of linear response theory, for which we have $\sigma_1=\chi_1 \phi_<(0)$,  $\sigma_2=\chi_2 \phi_{1>}(2)$, with $\chi_1$, $\chi_2$ are 2D susceptibilities. Generalising, $\chi_i=e^2\Pi^0_i$ for convenience, we obtain the solution of these equations for different coefficients, leading to

\begin{equation}
g(q,\omega) =  \frac{1}{D(q,\omega)}  \left\{
[q\epsilon_0(\epsilon_1-1)-\chi_1]
[q\epsilon_0(\epsilon_1+\epsilon_2)-\chi_2] -
[q\epsilon_0(\epsilon_1+1)+\chi_1]
[q\epsilon_0(\epsilon_1-\epsilon_2) +\chi_2]e^{-2dq}
\right\}
\label{gresp}
\end{equation}

\begin{equation}
D(q,\omega) \equiv 
[q\epsilon_0(\epsilon_1-1)-\chi_1]
[q\epsilon_0(\epsilon_1+\epsilon_2)-\chi_2] -
[q\epsilon_0(\epsilon_1-1)+\chi_1]
[q\epsilon_0(\epsilon_1-\epsilon_2) +\chi_2] e^{-2dq}\  ,
\label{grespdenom}
\end{equation}
where $\epsilon_1(\omega)$ is the dielectric function of the substrate between layers ``1" and ``2", $\chi_1$ and $\chi_2$ correspond to the susceptibilities of these two layers  and $d$ is the thickness of the substrate.  The plasmon dispersion equation is obtained from the solutions of $D(q,\omega)=0$ which we solve below. We note that when we set $\chi_2=0$ and take the limit $d\to \infty$, Eq.\ (\ref{eq2D}) yields the well established form \cite{BoPersson}

\begin{equation}
g_{2D}(q,\omega)=  1-\frac{1}{\frac{\epsilon_1+1}{2}-\frac{\chi_1}{2q\epsilon_0}}
\label{eq2D}
\end{equation}
which is the surface response function for a 2D layer embedded in a medium whose average background dielectric constant is $\epsilon_b=(\epsilon_1+1)/2$.  The plasma resonances, which Eq. \ (\ref{eq2D})  gives from its poles, are in agreement with the zeros of the dielectric function in Eq.\  (\ref{dielfn}).

\medskip
\par

\subsection{Damping rate}

We now turn to a critical issue in this paper which concerns the rate of damping of the plasmon modes by the single-particle excitations. If this rate of damping for a plasmon mode with frequency $\Omega_p$ is denoted as $\gamma$, then $D(\Omega_p+i\gamma,q)=0$ in the complex frequency space.  Carrying out a Taylor series expansion of both the real and imaginary parts, we have

\begin{eqnarray}
D(\Omega_p+i\gamma,q)&=& \text{Re}\ D(\Omega_p+i\gamma,q) +i \text{Im}\ D(\Omega_p+i\gamma,q)
\nonumber\\
&=& \texttt{Re} D(\Omega_p) +i\gamma \left.\frac{\partial}{\partial \Omega} \texttt{Re} D(\Omega)\right|_{\omega=\Omega_p}
+i\texttt{Im} D(\Omega_p) -   \gamma \left.\frac{\partial}{\partial \Omega} \texttt{Im} D(\Omega)\right|_{\omega=\Omega_p} +\cdots
\label{gammaX}
\end{eqnarray}
Therefore, setting the function in Eq.\ (\ref{gammaX}) equal to zero, we obtain $\gamma$ to lowest order as

\begin{equation}
\gamma= -  \frac{\texttt{Im} D(\Omega_p)}{\partial \texttt{Re} D(\omega)/\partial\omega|_{\Omega_{p}}}\ .
\label{gammaX1}
\end{equation}

With these formal results, we now evaluate the plasma spectra for double layer heterostructure. The expression shows the dependence of $\gamma $ on the imaginary part of $D(\Omega_p)$ and the Real part of $D(\omega)$ which in turn are dependent on the type of layer and the substrate considered. Eventually, we can infer that  the viability of plasmon modes can be tuned by the dielectric substrate thickness and by the choice of 2D layer. In addition, the rate of decay also helps us in maintaining the intensity and the frequency of the obtained plasmon mode. This could have great impact in the development of the quantum information sharing technology and the data storing devices.


\section{Numerical results and discussion}
\label{sec3}

\begin{figure}
\centering
\includegraphics[width=.65\textwidth]{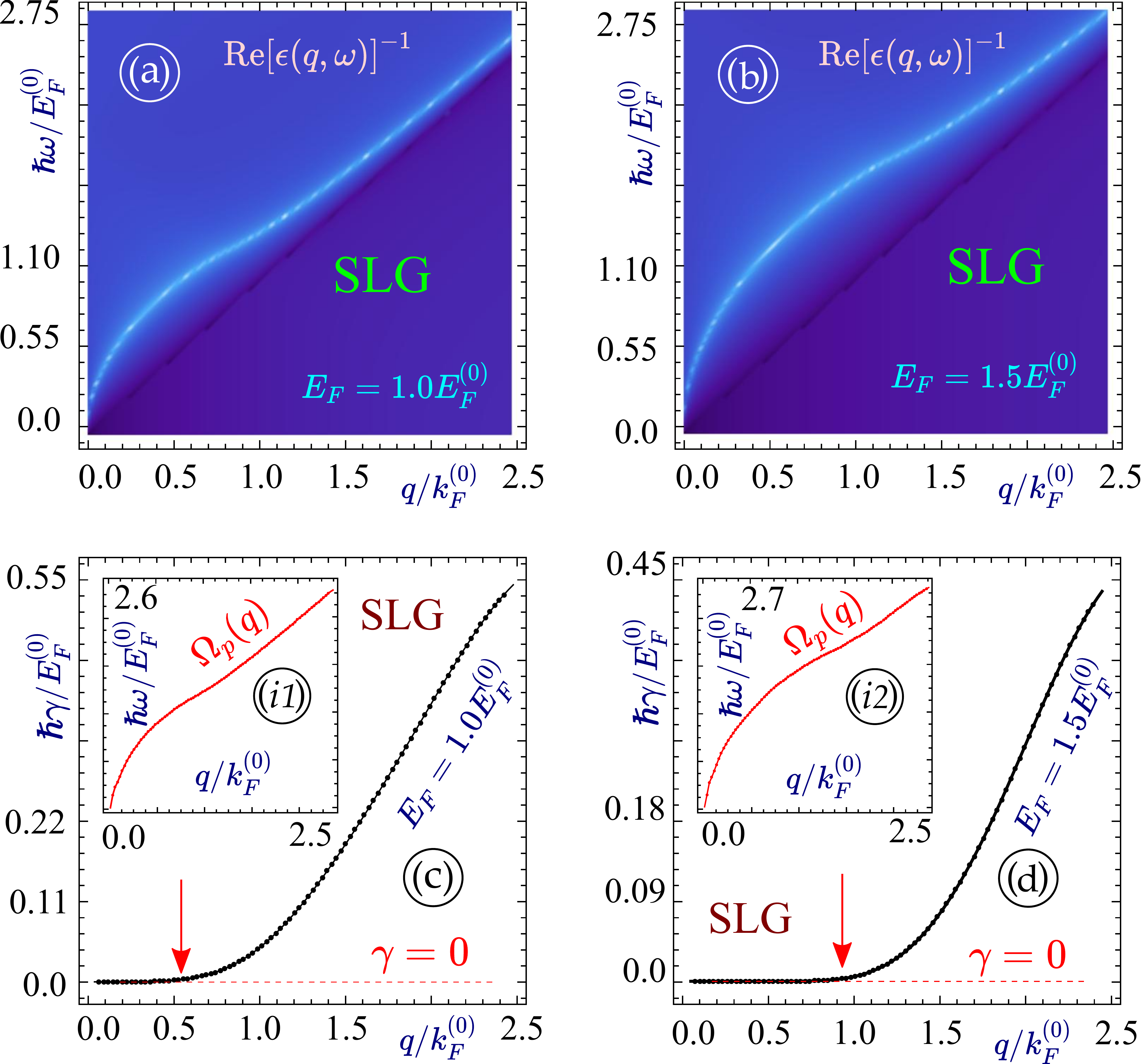}
\caption{(Color online) Plasmon frequency $\omega_p(q)$ and damping rates $\gamma(\omega_p(q),q)$ for an isolated graphene layer(SLG) with $E_F=1.0E_F^{(0)}$(left panels (a)and (c)) and $E_F=1.5E_F^{(0)}$(right panels (b) and(d)). Two top panels(a) and (b)demonstrate the plasmon dispersion (either damped or undamped obtained as $Re(\epsilon(q,\omega))=0$, the lower plots (c) and (d)describe the corresponding damping rate along the plasmon branches, also calculated and shown as insets(i1) and (i2).}
\label{Fig2}
\end{figure}
 
In our numerical calculations,  energy is scaled in units of $E_F^{(0)}$ and wave vector is scaled with $k_F^{(0)} =\sqrt{\pi n}$ which is in the experimental range for electron hole doping densities $n=10^{10}$  per cm$^2$. This gives $k_F^{(0)}=10^6$ per cm and $E_F^{(0)}$  is  equivalent to  $\sim$60meV.     
From the preceding discussion, in Sec.\  \ref{sec2}, it is clear that the plasmon mode for any system is given by the pole of the dielectric function obtained from Eq. \ref{grespdenom}. Thus, making use of it, we computed the plasmon mode dispersion  for a heterostructure based on graphene and silicene with/without a substrate. For this, we have first obtained a graphical result for graphene as shown in Fig.\  \ref{Fig2}. One can clearly see that a single branch plasmon mode originate from the origin in $q-\omega$ space which increases monotonically and decays out when the plasmon mode reaches the interband particle-hole excitation region. In Fig.\    \ref{Fig2}, the plasmon branches for two values of the Fermi energy are shown in panels (a) and (b). The damping rates of these plasmon modes are demonstrated in panel (c)  and (d), correspondingly, where it is distinctly shown by an arrow pointing at the boundary of the region where Landau damping takes place.  The rate of decay for both types of graphene are monotonically increasing signifying that the deeper into the single particle excitation region where the plasmon mode enters, the rate of plasmon decay becomes larger. That is the lifetime of the plasmon mode is decreased in the same manner.

 \medskip
\par

\begin{figure}
\centering
\includegraphics[width=.65\textwidth]{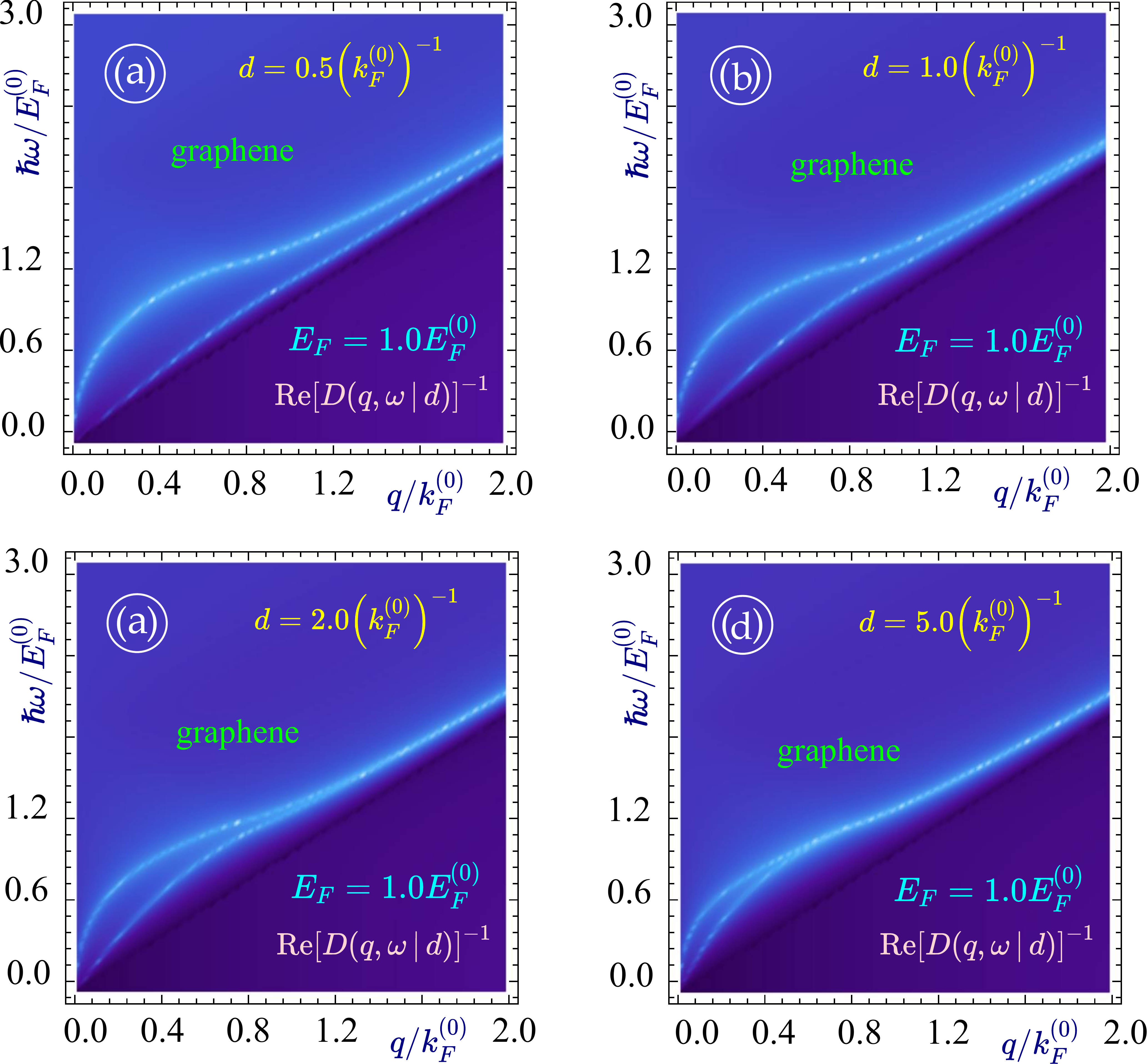}
\caption{(Color online)  Acoustic (lower) and optical (upper) plasmon modes for a pair of identical graphene layers with $E_F=1.0E_F^{(0)}$. Each panel corresponds to a different values of the separation between the layers corresponding to $d=0.5(k_F^{(0)})^{-1}$, $1.0(k_F^{(0)})^{-1}$, $2.0(k_F^{(0)})^{-1}$ and $5.0(k_F^{(0)})^{-1}$ as labeled. The plasmon dispersion   (either damped or undamped is obtained by solving
$Re(D(q,\omega|d))=0)$.}
\label{Fig3}
\end{figure}

\begin{figure}
\centering
\includegraphics[width=.65\textwidth]{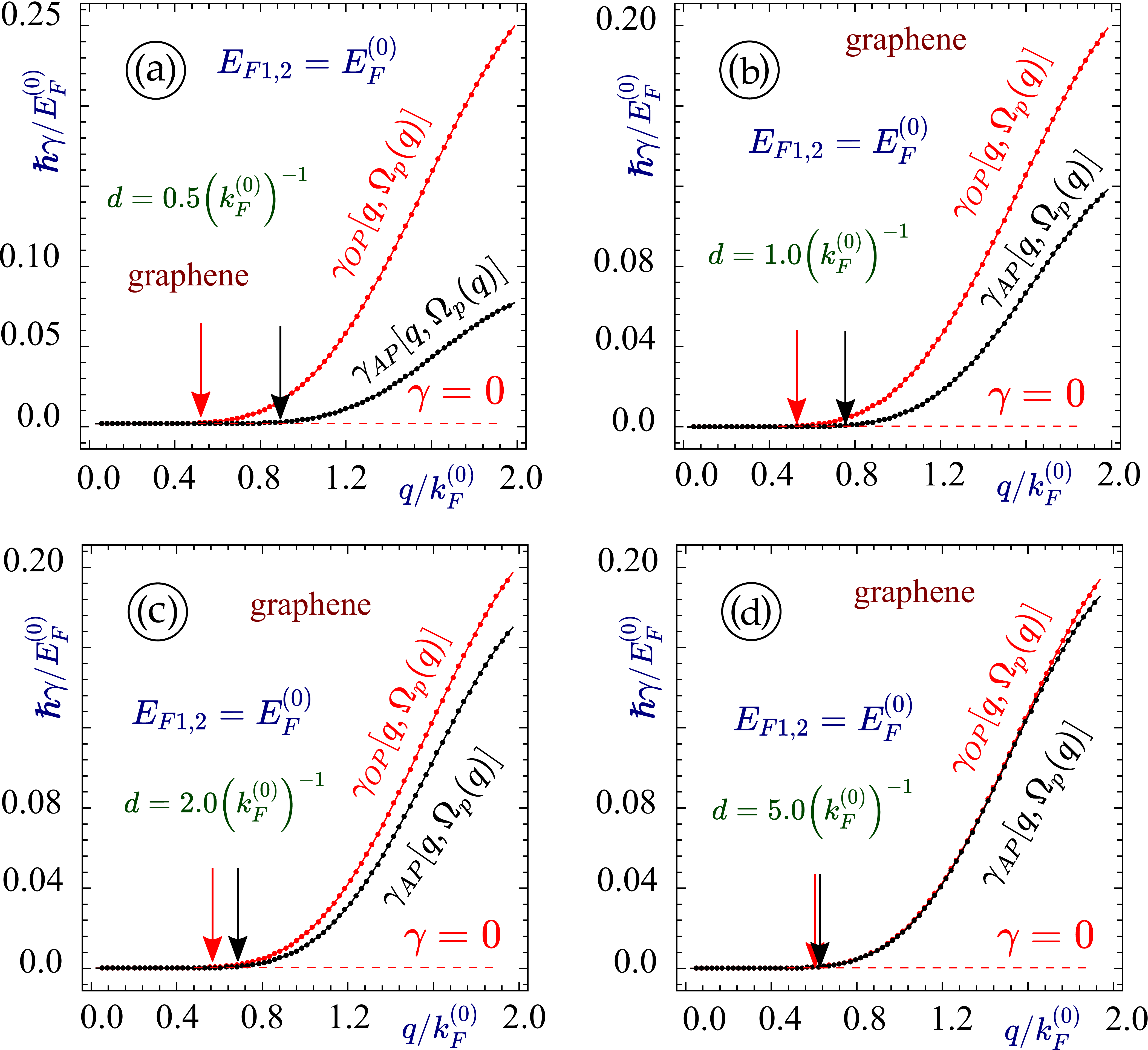}
\caption{(Color online) The damping rates corresponding to the acoustic and optical plasmon branches shown in Fig. 4 for two identical graphene layers with $E_F=1.0 E_F^{(0)}$. Each panel corresponds to a different values of the separation between the layers with $d=0.5(k_F^{(0)})^{-1}$, $1.0(k_F^{(0)})^{-1}$, $2.0(k_F^{(0)})^{-1}$ and $5.0(k_F^{(0)})^{-1}$ as labeled.}
\label{Fig4}
\end{figure}

Going next to the case when we have a structure with two graphene layers together, separated by various distances, a set of plots as shown in Fig.\  \ref{Fig3} are obtained with two branches of plasmon modes originating from the origin in the $q-\omega$ space. One can clearly see that the closer the graphene sheets are, the further apart are the plasmon modes. In Figs.  \ref{Fig3}(a), (b), (c) and (d), plasmon modes for a structure with two graphene layers separated by a distance of $0.5(k_F^{(0)})^{-1}$, $1.0(k_F^{(0)})^{-1}$, $2.0(k_F^{(0)})^{-1}$ and $5.0(k_F^{(0)})^{-1}$ are shown, respectively,  and all the plots portrays that the further apart the graphene layers are, the closer the two plasmon branches become. For this same set of figures with the other parameters remaining the same, the plasmon decay rate is shown in Fig.\  \ref{Fig4} which shows that the plasmon decay for the lower plasmon branch always starts at larger wave vector value  in comparison  to upper plasmon branch. As the distance of separation is increased, the two plasmon branches come closer and their decay also starts from the same value of wave vector and the rate of decay is almost same in value.

\medskip
\par
 
\begin{figure}
\centering
\includegraphics[width=.65\textwidth]{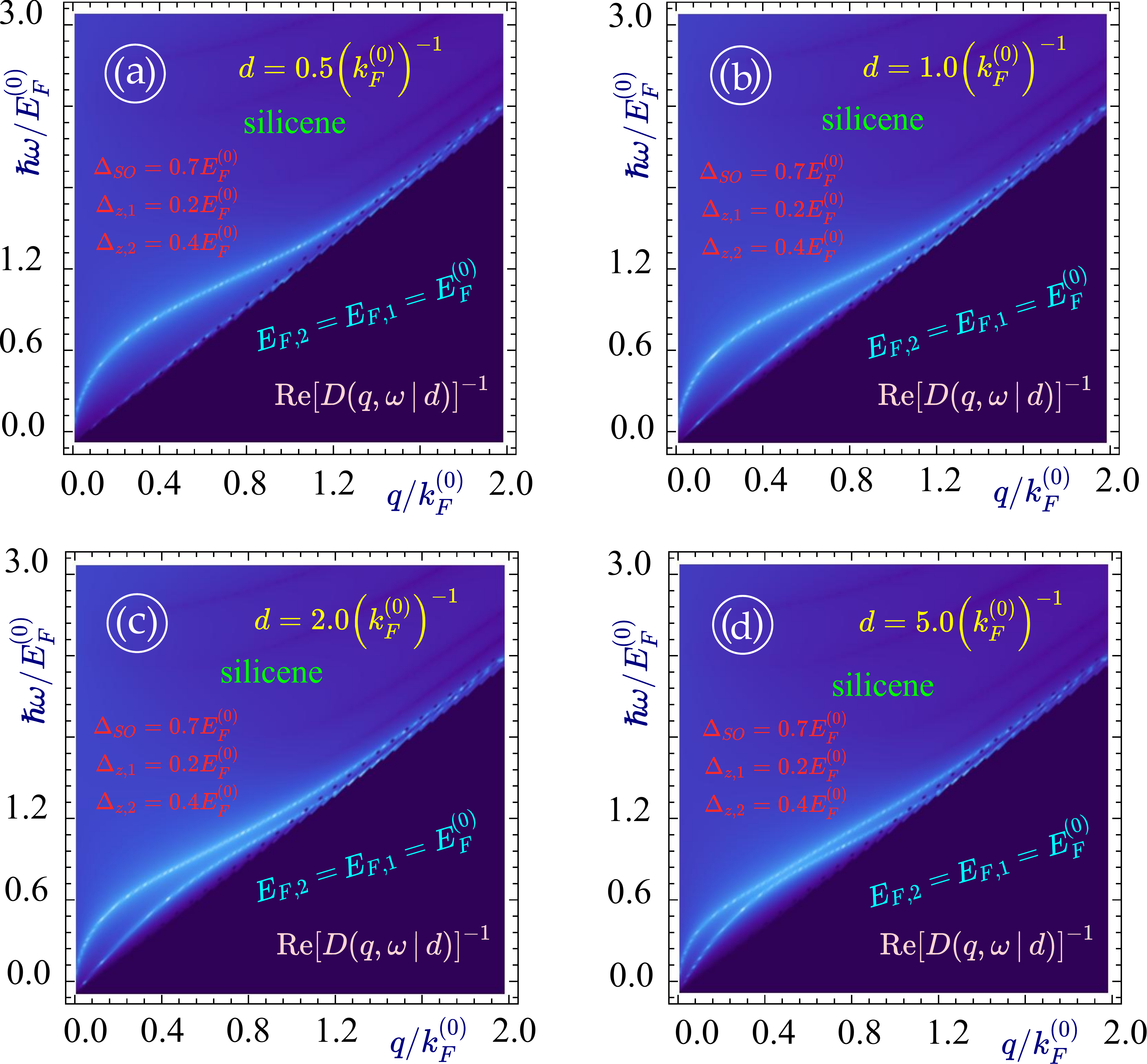}
\caption{(Color online) Acoustic  (lower) and optical  (upper) plasmon dispersions for two silicene layers with $E_F=1.0 E_F^{(0)}$ and the band gaps $\Delta_{SO,1}=\Delta_{SO,2}=0.7E_F^{(0)}$, $\Delta_{z,1}=0.2E_F^{(0)}$ and $\Delta_{z,2}=0.4E_F^{(0)}$. Each panel corresponds to a different values of the separation between the layers $d=0.5(k_F^{(0)})^{-1}$, $1.0(k_F^{(0)})^{-1}$, $2.0(k_F^{(0)})^{-1}$ and $5.0(k_F^{(0)})^{-1}$ as labeled.  }
\label{Fig5}
\end{figure}

Now, in addition, we carried out an investigation of the plasmon modes and their decay rate for the structure with two silicene layer for various  separations. The graphical results for these calculations are shown in Fig.\  \ref{Fig5} where we again have two plasmon modes originating from the origin of $q-\omega$ plane.  As in the case for a two-graphene-layer structure, we again notice a similar effect on two plasmon branches coming closer to each other when their   separation  increases. This clearly is demonstrated in Figs.\  \ref{Fig5}(a), (b), (c) and (d)  for their distance apart of   $0.5(k_F^{(0)})^{-1}$, $1.0(k_F^{(0)})^{-1}$, $2.0(k_F^{(0)})^{-1}$, and $5.0(k_F^{(0)})^{-1}$ respectively.

\medskip
\par
\begin{figure}
\centering
\includegraphics[width=.65\textwidth]{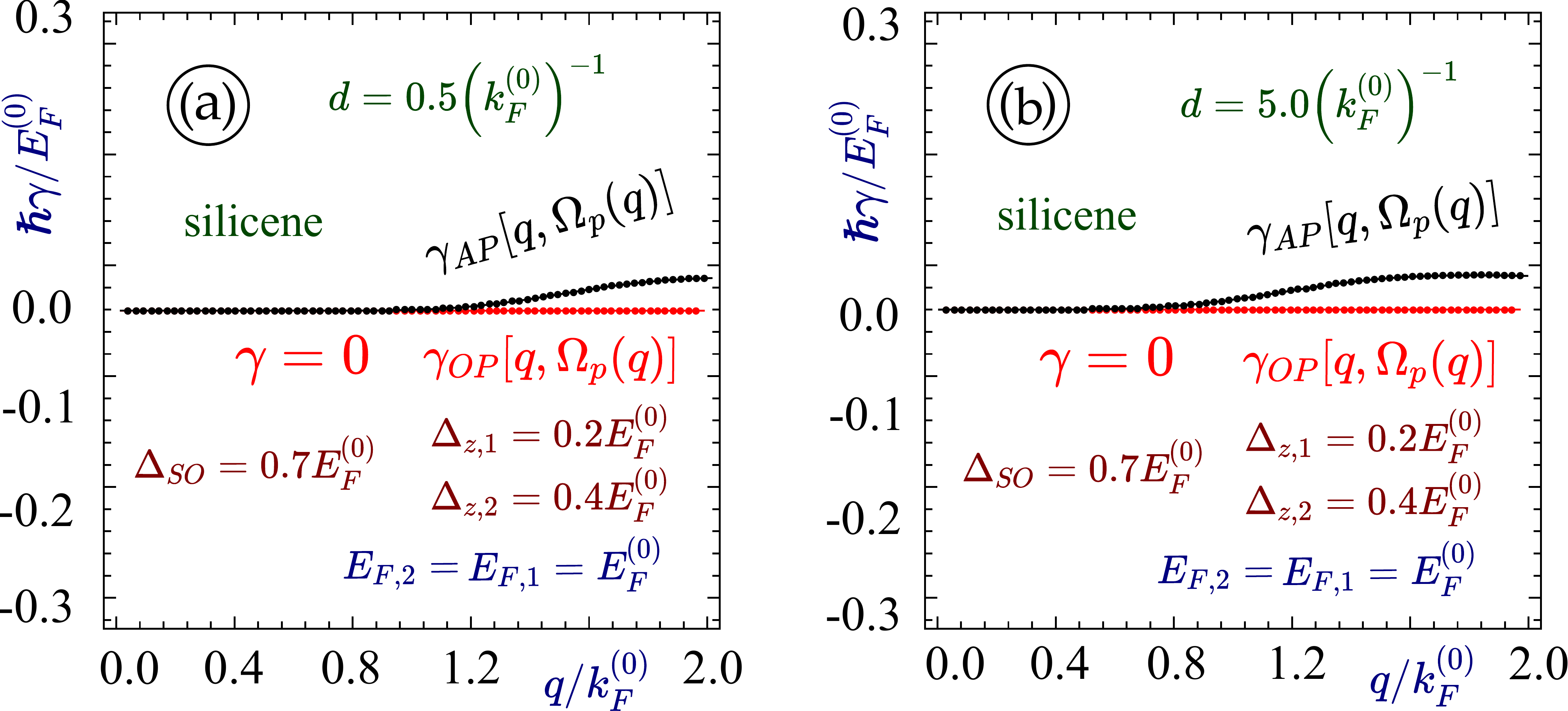}
\caption{(Color online) The damping rates corresponding to the acoustic and optical plasmon branches calculated in Fig. 6 for two silicene layers with $E_F=1.0E_F^{(0)}$, and the band gaps $\Delta_{SO,1}=\Delta_{SO,2}=0.7E_F^{(0)}$, $\Delta_{z,1}=0.2E_F^{(0)}$ and $\Delta_{z,2}=0.4E_F^{(0)}$.   Each panel corresponds to a different values of the separation between  layers with $d=0.5(k_F^{(0)})^{-1}$,  and $5.0(k_F^{(0)})^{-1}$ as labeled.}
\label{Fig6}
\end{figure}

  As a representative calculation, we investigated the decay rate of plasmon modes for silicene-silicene structure when their separation is $d=0.5(k_F^{(0)})^{-1}$ and  $d=5.0(k_F^{(0)})^{-1}$. Figure \ref{Fig6} shows that the upper plasmon mode does not decay at all and the lower plasmon branch decays after reaching a critical wave vector. This behavior is  due to the presence of a band gap for silicene,  resulting in an opening  in the single particle excitation region,  which provides a larger area in $q-\omega$ space  for the plasmon mode to survive.  The upper plasmon branch in this case has a larger space and is more likely  to self-sustain for a longer period of time without damping. On the other hand,  the lower plasmon branch  enters the intraband single-particle excitation region where it decays. The rate of decay starts from a critical  value of the wave vector and  the magnitude of this decay rate monotonically increases.

\medskip
\par

\begin{figure}
\centering
\includegraphics[width=.65\textwidth]{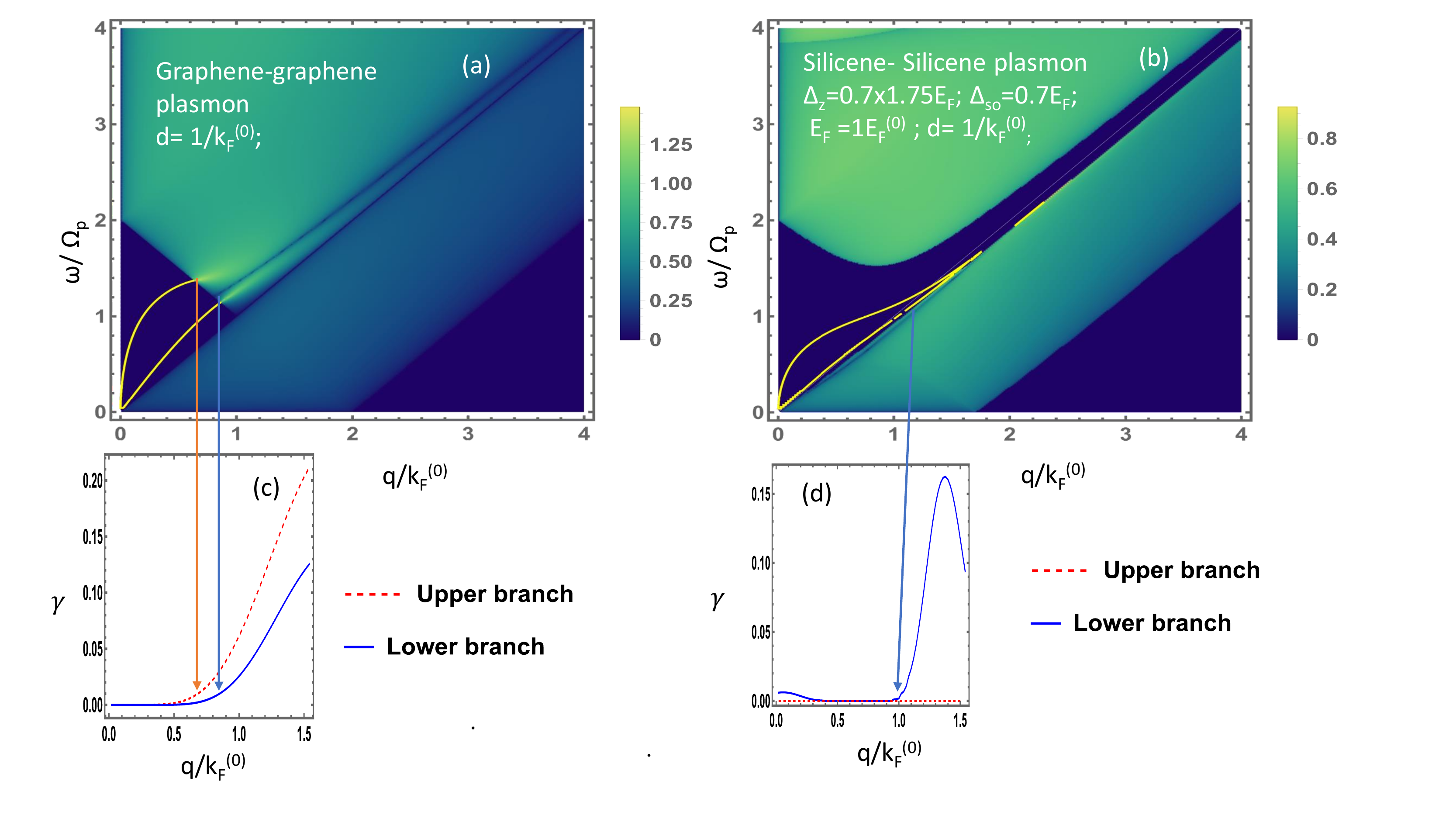}
\caption{(Color online)  Plasmon modes for (a) graphene-graphene structure and (b) silicene-silicene structure with corresponding plasmon damping rate in Fig (c) and (d), respectively.}
\label{Fig7}
\end{figure}

A comparison of plasmon modes and their decay for graphene-graphene and silicene-silicene structures is shown along with the single-particle excitation regions in Fig.\ \ref{Fig7}. The figure in panel (a) of Fig.\ref{Fig7} shows that two plasmon modes which originate from the origin of the frequency-momentum space is steadily increased but decays after it reaches the boundary of  the single-particle excitation region. Corresponding red lines are drawn to further clarify the point where the actual decay begins. The dark triangular region is the area where the plasmon mode survives without Landau damping and mathematically, in this region, the imaginary part of the polarization function of graphene is zero. This means that  the plasmon mode has self-sustaining oscillations.  The green region where the imaginary part of the polarization function is nonzero is the single particle excitation region where  the plasmon mode decays into particle hole mode. The corresponding decay rate figure  below this panel  shows that the rate of decay for the upper plasmon branch is greater and its critical wave vector is smaller compared with the lower plasmon branch.

 \medskip
\par
 
 Similar plots for silicene-silicene structures were demonstrate figure \ref{Fig7} (b) where one can clearly see the opening of a gap in the single-particle excitation region yielding two parts which is a significant effect arising from band gap.  The imaginary part of the polarization function in this gap region is zero where the plasmon mode can sustain its oscillation for long time. The upper break away region is single-particle excitation region due to interband  transitions of electrons  from the valence to the conduction band and the lower break away region is  the intraband single particle excitations region which is due to   transistions within the same  band from below to above the Fermi level.   In Fig.\   . \ref{Fig7}, two plasmon modes originate from the origin as demonstrated in the figure. The upper plasmon mode survives without damping  over a wider   range of wave vector and the plasmon branch enters the gap created  by the opening within the single-particle excitation region.  The corresponding decay rate appearing below the plasmon dispersion shows that the upper plasmon branch does  not decay at all. Whereas  for the case of the  lower plasmon branch, the plasmon mode near the origin lies closer to the intraband single-particle region.   Consequently, there is small plasmon decay rate as illustrated in the corresponding figure in the  panel fig \ref{Fig7}(d) below. As the plasmon mode  rises, it gets separated from the single-particle excitation region where the decay rate is zero and as it moves further away from the origin the plasmon branch comes in contact with the single particle excitation region where we notice the Landau damping again. Correspondingly, the decay rate is increased monotonically,  reaches a maximum before drops down, indicating the reappearance of an undamped plasmon branch  at a larger value of wave vector. Another noticeable effect observed here is the closeness of the plasmon branches and the plasmon decay rate which can be altered by changing the layer separation, this effect may be used as another plasmon mode tuning parameter.
\medskip
\par

\begin{figure}
\centering
\includegraphics[width=.65\textwidth]{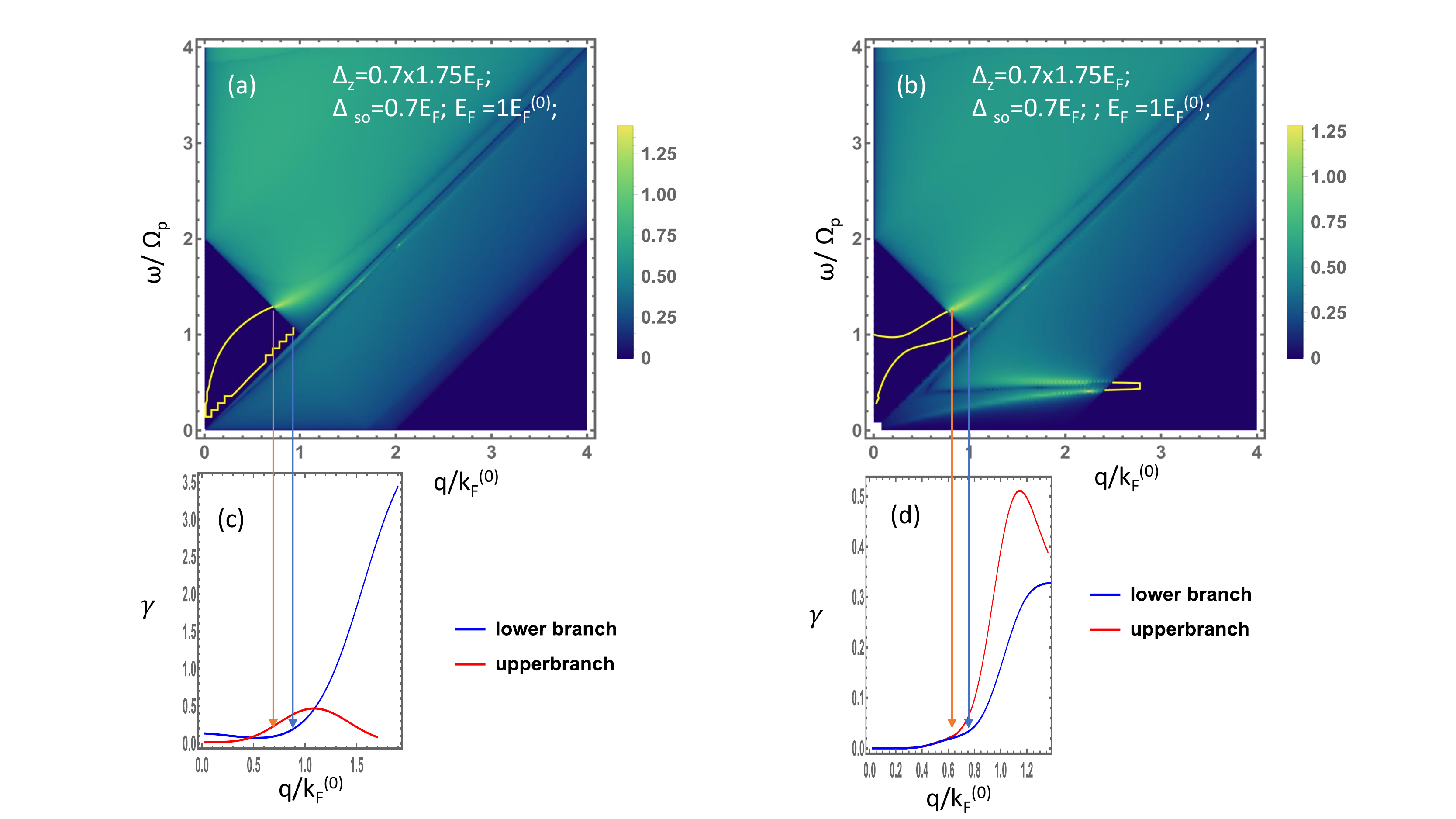}
\caption{(Color online) Plasmon mode dispersion for two layer of of 2D materials separated by a distance of  $2 k_F^{-1}$. In (a) Both layers are silicene, $\epsilon_1(\omega)=1-\Omega_p^2/\omega^2$ and $\epsilon_2=1$ for the substrate.   (b) The same as (a) except that the  silicene layers are replaced by graphene.}
\label{Fig8}
\end{figure}

In order to extract more information about the plasmonic behavior, in Fig.\ \ref{Fig8} (a), we have presented the figure to show the result  highlighting the changes in the plasmonic nature for a structure with  different types of layer and a substrate. In Fig.\  \ref{Fig8}(a), we demonstrate the plasmon mode for a structure with silicene and graphene separated by a distance of $1.0(k_F^{(0)})^{-1}$ with vacuum in between. One can clearly observe two plasmon modes originate from the  origin in $q-\omega$ space.   A special effect of overcoming the single particle excitation region of silicene by the single-particle excitation region of graphene is observed which causes the  shortening of the lower plasmon branch which used to be there for silicene-silicene structure. AS soon as the plasmon branch reaches this region, the plasmon mode decays into particle hole mode just because of the replacine one silicene layer by a graphene layer in silicene-silicene structure. The effect due to the band gap in silicene is just nullified. Furthermore, the result of adding a substrate between the silicene and graphene layer is illustrated in panel (b) of Fig. \ref{Fig8}.  In this case, we could see a new plasmon branch originating from the bulk plasma frequency and one plasmon branch originating from the origin. Here, due to the presence of a substrate,  the lower plasmon branch bend sharply towards the intraband  single-particle excitation region where it decays causing complete disappearance.
  The upper plasmon branch and the plasmon from the bulk plasmon frequency become closer and move towards the interband single-particle excitation region where they get damped. 
\medskip
\par

These new effects on the plasmon branches in this type of structure was not reported previously.  Results of this type are helpful in developing  electronic and quantum computing devices where  knowledge of plasmonic behavior of material is very essential.

\medskip
\par

\section{Concluding Remarks}
\label{sec4}

 In summary,   we   have shown that the effect due to the addition of a substrate and the difference from introducing different types of 2D materials in the heterostructure resulting a novel tunning technique of the plasmon excitations associated with these 2D systems.  A completely new effect of  a plasmon branch emerging from the bulk plasmon frequency is seen which would be very helpful in engineering  computing devices.   Another discovery is the  disappearance of the lower plasmon branch and the suppression of the silicene band gap effect. These  are other interesting new effects seen from our calculations. We have also developed an approach for calculating the decay rates for the plasmons  due to Landau damping by the particle-hole  modes. 

\medskip
\par
Additionally, our results infer that the number of plasmon branches emerging from the origin can be varied by choosing the number of 2D layer. In brief, we can say that our study gives an important idea about the plasmonic behavior of a graphene-silicene based heterostructure which would be very helpful in carrying out further study of other type of heterostructure including various low dimensional material.

\section*{Acknowledgement(s)}

G.G. would like to acknowledge the support from the Air Force Research Laboratory (AFRL)
through Grant No. FA9453-21-1-0046

\end{document}